\pdfoutput=1

\PassOptionsToPackage{table}{xcolor}
\documentclass[conference]{IEEEtran}
\IEEEoverridecommandlockouts
\usepackage{xspace}
\usepackage{url}
\usepackage{bm}
\usepackage{booktabs}
\usepackage{tabularx}
\usepackage{hhline}
\usepackage[table]{xcolor}
\usepackage{multirow}
\usepackage{subcaption}
\usepackage{wrapfig}
\usepackage{array}
\newcolumntype{L}[1]{>{\raggedright\let\newline\\\arraybackslash\hspace{0pt}}m{#1}}
\newcolumntype{C}[1]{>{\centering\let\newline\\\arraybackslash\hspace{0pt}}m{#1}}
\newcolumntype{R}[1]{>{\raggedleft\let\newline\\\arraybackslash\hspace{0pt}}m{#1}}
\newcommand{\bigo}{\mathcal{O}}
\usepackage{perpage}
\MakePerPage{footnote}
\usepackage{graphicx}
\usepackage{listings}
\usepackage{xcolor}
\usepackage{float}
\newfloat{lstfloat}{htbp}{lop}
\floatname{lstfloat}{Listing}

\definecolor{commentsColor}{rgb}{0.497495, 0.497587, 0.497464}
\definecolor{keywordsColor}{rgb}{0.000000, 0.000000, 0.635294}
\definecolor{stringColor}{rgb}{0.558215, 0.000000, 0.135316}
\lstdefinestyle{mystyle}{
    captionpos=b,
    breaklines=true,
    tabsize=2,
    frame=b,
    showstringspaces=false,
    numberstyle=\tiny\color{commentsColor},
    rulecolor=\color{black},
    commentstyle=\color{commentsColor}\textit,
    stringstyle=\color{stringColor},
    keywordstyle=\color{keywordsColor}\bfseries,
    basicstyle=\ttfamily\scriptsize,
}
\usepackage{microtype}

\def\BibTeX{{\rm B\kern-.05em{\sc i\kern-.025em b}\kern-.08em
    T\kern-.1667em\lower.7ex\hbox{E}\kern-.125emX}}

\newcommand{\cg}{CompilerGym\xspace}
\newcommand{\pg}{\textsc{ProGraML}\xspace}
\newcommand{\llvm}{LLVM\xspace}
\newcommand{\gcc}{GCC\xspace}

\usepackage{hyperref}
\hypersetup{colorlinks=false}

\begin{document}

\title{CompilerGym: Robust, Performant Compiler Optimization Environments for AI Research}

\author{%
    \IEEEauthorblockN{%
        Chris Cummins,\hspace{.5em} 
        Bram Wasti,\hspace{.5em} %
        Jiadong Guo,\hspace{.5em} %
        Brandon Cui,\hspace{.5em} %
        Jason Ansel,\hspace{.5em} %
        Sahir Gomez,\\ %
        Somya Jain,\hspace{.5em} %
        Jia Liu,\hspace{.5em} %
        Olivier Teytaud,\hspace{.5em} %
        Benoit Steiner,\hspace{.5em} %
        Yuandong Tian,\hspace{.5em} %
        Hugh Leather%
    }
    \IEEEauthorblockA{%
        \textit{Facebook}\\
        cummins@fb.com
    }
}

\maketitle

\begin{abstract}
Interest in applying Artificial Intelligence (AI) techniques to compiler
optimizations is increasing rapidly, but compiler research has a high entry
barrier. Unlike in other domains, compiler and AI researchers do not have access
to the datasets and frameworks that enable fast iteration and development of
ideas, and getting started requires a significant engineering investment. What
is needed is an easy, reusable experimental infrastructure for real world
compiler optimization tasks that can serve as a common benchmark for comparing
techniques, and as a platform to accelerate progress in the field.

We introduce \cg{}\footnote{Available at: \url{https://compilergym.ai}}, a set
of environments for real world compiler optimization tasks, and a toolkit for
exposing new optimization tasks to compiler researchers. \cg{} enables anyone to
experiment on production compiler optimization problems through an easy-to-use
package, regardless of their experience with compilers. We build upon the
popular OpenAI Gym interface enabling researchers to interact with compilers
using Python and a familiar API.

We describe the \cg{} architecture and implementation, characterize the
optimization spaces and computational efficiencies of three included compiler
environments, and provide extensive empirical evaluations. Compared to prior
works, \cg{} offers larger datasets and optimization spaces, is 27$\times$ more
computationally efficient, is fault-tolerant, and capable of detecting
reproducibility bugs in the underlying compilers.

In making it easy for anyone to experiment with compilers -- irrespective of
their background -- we aim to accelerate progress in the AI and compiler
research domains.

\end{abstract}

\section{Introduction}

There is a growing body of work that shows how the performance and portability
of compiler optimizations can be improved through
autotuning~\cite{autotuning-survey}, machine learning~\cite{ml4sysreview}, and
reinforcement learning~\cite{mlgo,autophase,neurovectorizer}. The goal of these
approaches is to supplement or replace the optimization decisions made by
hand-crafted heuristics with decisions derived from empirical data. Autotuning
makes these decisions by automatically searching over a space of configurations.
This is effective, but search may be prohibitively costly for large search
spaces, and must be repeated from scratch for each new problem instance. The
promise of supervised and reinforcement learning techniques is to reduce or
completely eliminate this search cost by inferring optimization decisions from
patterns observed in past data.

\begin{figure}[t]
  \centering %
  \includegraphics[width=.9\linewidth]{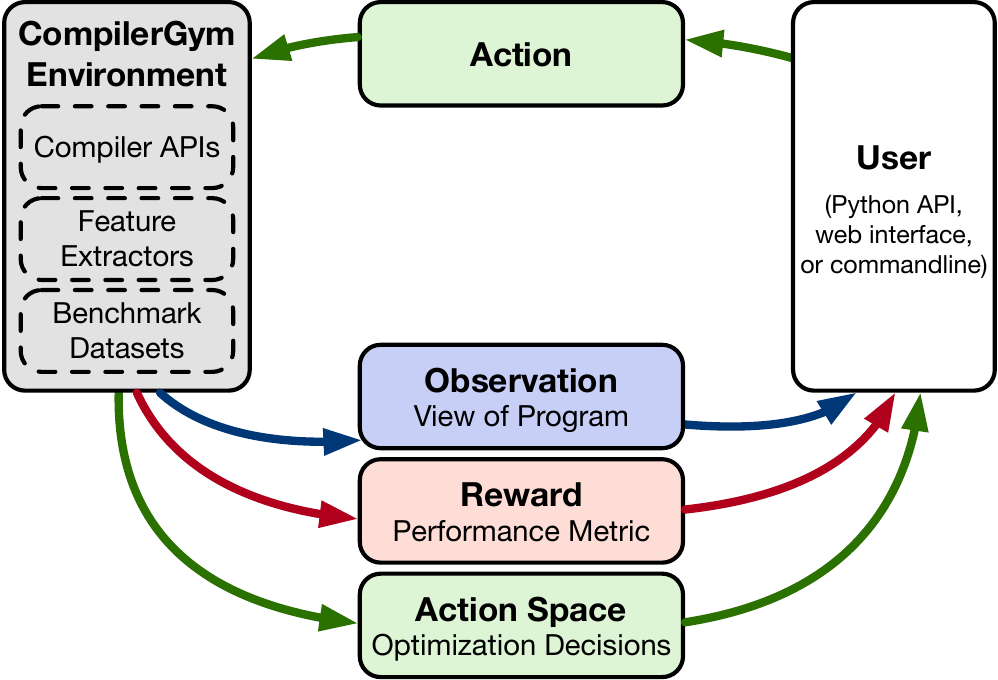}
  \caption{%
    The \cg{} interaction loop. A \cg{} environment exposes an observation,
    reward, and action space. The user's goal is to select the action that will
    lead to the greatest cumulative reward. This may be through hand-crafted
    heuristics, search, supervised machine learning, or reinforcement learning.%
    \vspace{-.5em}%
  }%
  \label{figure:rl-loop}%
\end{figure}

Despite many strong experimental results showing that these techniques
outperform human experts~\cite{ml4sysreview,autotuning-survey,big-code-survey},
the complexity of experimental infrastructure for compiler research hampers
progress in the field. In many other fields there are simple environments, each
using standard APIs that machine learning researchers can interact with. From
Atari games to physics simulations, a known interface abstracts the problems to
the point that AI researchers do not need deep knowledge of the problem to apply
their machine learning techniques. \cg provides just that for compilers. AI
researchers can solve compiler problems without being compiler experts, and
compiler experts can integrate state-of-the-art ML without being AI experts.

To support this ease of use and performance \cg offers the following key
features:
\begin{enumerate}
\item \textbf{Easy to install.} Precompiled binaries for Linux and macOS can be
  installed with a single command.
\item \textbf{Easy to use.} Builds on the Gym~\cite{gym} API that is easy to
  learn and widely used by researchers.
\item \textbf{Comprehensive.} Includes a full suite of millions of benchmarks.
  Provides multiple kinds of pre-computed program representations and
  appropriate optimization targets and reward functions out of the box.
\item \textbf{Reproducible.} Provides validation for correctness of results and
public leaderboards to aggregate results.
\item \textbf{Accessible.} Includes code-free ways to explore \cg{}
environments, such as an interactive command line shell and a browser-based
graphical user interface.
\item \textbf{Performant.} Supports the high throughput required for large-scale
experiments on massive datasets.
\item \textbf{Fault-tolerant.} Detects and gracefully recovers from flaky
compiler errors that can occur during autotuning.
\item \textbf{Extensible.} Removes the substantial engineering effort required
  to expose new compiler problems for research and integrate new machine
  learning techniques.
\end{enumerate}

\noindent
In this paper, we make the following contributions:
\begin{itemize}
  \item We introduce \cg{}, a Python library that formulates compiler
    optimization problems as easy-to-use Gym~\cite{gym} environments with a
    simple API.
  \item We provide environments for three compiler optimization problems: LLVM
    phase ordering, GCC flag selection, and CUDA loop nest generation. The
    environments are designed from the ground up for large-scale
    experimentation: they are 27$\times$ faster than prior works, expose larger
    search spaces, include millions of programs for training, and support
    optimizing for both code size and runtime.
  \item We demonstrate the utility of \cg{} as a platform for research by
    evaluating a multitude of autotuning and reinforcement learning techniques.
    By using a standard interface, \cg{} seamlessly integrates with third party
    libraries, offering a substantial reduction in the engineering effort
    required to create compiler experiments.
  \item We release a suite of tools to lower the barrier-to-entry to compiler
    optimization research: the core \cg{} library and environments, a toolkit
    for integrating new compiler optimization problems, public leaderboards to
    aggregate and verify research results, a web interface and API, extensive
    command line tools, and large offline datasets comprising millions of
    performance results.
\end{itemize}

\begin{lstfloat}[t]
\begin{lstlisting}[%
  language=Python,%
  caption={%
    Example of the \cg{} environment API. A \cg environment  builds on
    \texttt{\small gym.Env}, formulating a compiler optimization task as a
    Markov Decision Process, and provides additional compiler-specific
    functionality such as saving the compiled program to disk.%
    \vspace{-.5em}%
  },%
  label=listing:user-api-demo
]
import compiler_gym
# Create a new environment, selecting the compiler to
# use, the program to compile, the feature vector to
# represent program states, and the optimization target:
env = compiler_gym.make(
    "llvm-v0",
    benchmark="cbench-v1/qsort",
    observation_space="Autophase",
    reward_space="IrInstructionCount",
)
# Start a new compilation session:
observation = env.reset()
# Run a thousand random optimizations. Each step of the
# environment produces a new state observation and reward:
for _ in range(1000):
    observation, reward, done, info = env.step(
        env.action_space.sample()  # User selects action.
    )
    if done:
        env.reset()
# Save output program:
env.write_bitcode("/tmp/output.bc")
\end{lstlisting}
\end{lstfloat}

\begin{figure}[t]
  \centering %
  \includegraphics[width=.9\linewidth]{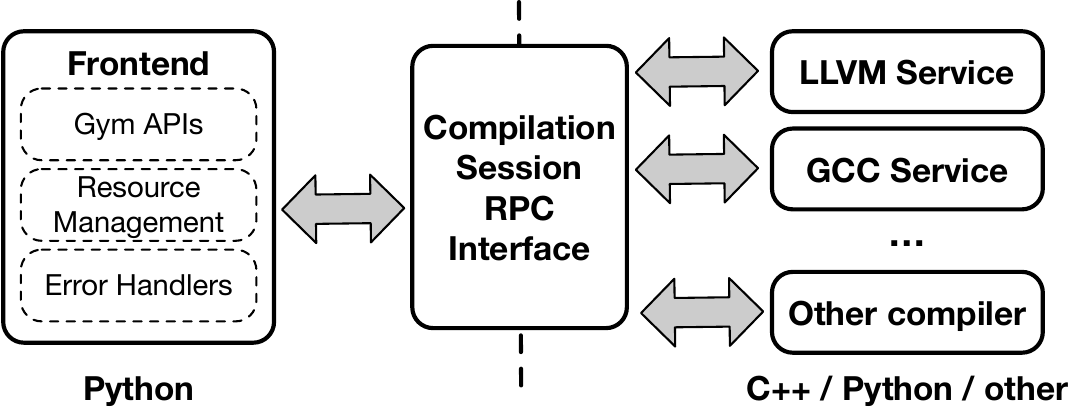}
  \caption{%
    The client-server architecture. The dashed line indicates the boundary
    between the frontend Python process that the user interacts with and the
    backend compiler services. The processes communicate over RPC, providing
    simple distribution, parallelization, and fault tolerance.%
    \vspace{-.5em}%
  }%
  \label{figure:client-service}%
\end{figure}

\section{System Architecture}

\cg{}'s architecture comprises two components: a Python frontend that implements
the Gym APIs and other user-facing tools, and a backend that provides the
integrations with specific compilers.

\subsubsection*{Frontend} The \cg{} frontend is a Python library that exposes
compiler optimization tasks using the OpenAI Gym~\cite{gym} environment
interface. Figure~\ref{figure:rl-loop} shows the interaction loop for the Gym
environments. This allows researchers to interact with important compiler
optimization problems in a familiar language and vocabulary with which many are
comfortable. The frontend is described in Section~\ref{section:frontend}.

\subsubsection*{Backend} \cg{} uses a client-server architecture, shown in
Figure~\ref{figure:client-service}. This design provides separation of concerns
as systems developers can easily add support for new compiler problems by
implementing a simple Compilation Session interface that comprises only four
methods. The backend is described in Section~\ref{section:backend}.

\section{Frontend API and Tools}%
\label{section:frontend}

This section describes \cg{}'s user-facing tools. We first describe the core
formulation of compiler optimization problems as Gym environments, then the API
extensions and other features tailored for compiler optimization research.

\subsection{OpenAI Gym Environments}

We formulate compiler optimization tasks as Markov Decision Processes (MDPs) and
expose them as environments using the popular OpenAI Gym~\cite{gym} interface. A
Gym environment comprises five ingredients:

1) An \emph{Action Space} defines the set of possible actions that can be taken
from a given MDP state. In \cg{}, action spaces can be composed of discrete
choices (e.g. selecting an optimization pass from a finite set), continuous
choices (e.g. selecting a function inlining threshold), or any combination of
the two. The action space can change between states, such as in the case where
one optimization precludes another.

2) An \emph{Observation Space} from which observations of the MDP state are
drawn. \cg{} environments support multiple observation types such as numeric
feature vectors generated by compiler analyses, control flow graphs, and strings
of compiler IR. Each environment exposes multiple observation spaces that can be
selected from or composed.

3) A \emph{Reward Space} defines the range of values generated by the reward
function, used to provide feedback on the quality of a chosen action, either
positive or negative. In \cg{}, reward spaces can be nondeterministic (e.g.
change in program runtime), platform specific (e.g. change in the size of a
compiled binary), or entirely deterministic.

4) A \emph{Step} operator applies an action at the current state and responds
with a new observation, a reward, and a signal that indicates whether the MDP
has reached a terminal state. Not all compiler optimization problems have
terminal states.

5) A \emph{Reset} operator resets the environment to an initial state and
returns an initial observation.

Listing~\ref{listing:user-api-demo} demonstrates how the core \cg{} API is used.
A \texttt{\small make()} function instantiates a subclass of the \texttt{\small
gym.Env} environment that represents a particular compiler optimization task.
The Gym interface is self describing: the action space and observation spaces
are described by \texttt{\small action\_space} and \texttt{\small
observation\_space} attributes, respectively. This enables \cg{} environments to
be integrated directly with techniques that are compatible with other Gym
environments. Listing~\ref{listing:rllib-demo} shows one such integration.

In interacting with an environment the user's goal is to select the sequence of
actions that maximizes the cumulative reward. Although Gym is designed primarily
for reinforcement learning research, it makes no assumptions about the structure
of user code and therefore can be used with a wide range of approaches. For a
single environment, the best sequence of actions may be found through search. To
generalize a solution that works for unseen environments, a \emph{policy} is
learned to map from observation to optimal actions, or a \emph{$Q$-function} is
learned to give expected cumulative rewards for state-action pairs.

\subsection{API Extensions for Compiler Optimization}%
\label{subsection:api-extensions}

The advantage of the Gym interface is that it is simple and can be used across a
range of domains. We supplement this interface with additional APIs that are
specific to compilers.

\subsubsection{Benchmark Datasets}

An instance of a compiler optimization environment requires a program to
optimize. We refer to these programs as \emph{benchmarks}, and collections of
benchmarks as \emph{datasets}. We designed an API to manage datasets that
efficiently scales to millions of benchmarks, and a mechanism for downloading
datasets from public servers. This API supports program generators (like
CSmith~\cite{csmith}), compiling user-supplied code to use as benchmarks,
iterating and looping over sets of benchmarks, and specifying an input dataset
and execution environment for running compiled binaries.

\subsubsection{State Serialization}

We provide a mechanism to save and restore environment state that includes the
benchmark, action history, and cumulative reward.

\subsubsection{Validating States}

Serialized states can be replayed to validate that results are reproducible. We
use this to ensure reproducibility of the underlying compiler infrastructure.
For example, we detected a nondeterminism bug in an LLVM optimization
pass\footnote{LLVM's \texttt{-gvn-sink} pass contains an operation that sorts a
vector of basic block pointers by address, causing inconsistent output.}; we
removed this pass from \cg{}.

\begin{lstfloat}[t]
\begin{lstlisting}[%
    language=Python,%
    caption={%
        Using RLlib~\cite{rllib} to train a reinforcement learning agent on one
        of the \cg{} environments.%
        \vspace{-.5em}%
    },%
    label=listing:rllib-demo
]
import compiler_gym
from ray import tune
from ray.rllib.agents.ppo import PPOTrainer

def make_env(config):
  # Create an LLVM environment using the Autophase
  # observation space and instruction count rewards.
  env = compiler_gym.make("llvm-autophase-ic-v0")
  # Optionally create a time limit for the RL agent.
  env = compiler_gym.wrappers.TimeLimit(env, 45)
  # Loop over the NPB benchmark suite for training.
  dataset = env.datasets["benchmark://npb-v0"]
  env = compiler_gym.wrappers.CycleOverBenchmarks(
      env, dataset.benchmarks()
  )
  return env

tune.register_env("CompilerGym", make_env)
tune.run(PPOTrainer, config={"env": "CompilerGym"})
\end{lstlisting}
\end{lstfloat}

\subsubsection{Validating Semantics}

For runnable benchmarks, we provide an additional layer of results validation
that automatically applies a differential testing~\cite{difftesting} regime to
detect correctness errors in the compiled binaries. For the LLVM environments we
also integrate LLVM's address, thread, and undefined behavior sanitizers to
detect program logic errors.

\subsubsection{Lazy and batched operations}

Typically, the observation and reward spaces of a Gym environment are determined
at construction time, and each \texttt{\small step()} operation takes a single
action and produces a single observation and reward. We extend this method in
\cg{} environments to optionally accept multiple actions, and a list of
observation and reward spaces to compute and return. Passing multiple actions
enables the backend to more efficiently execute them in a single batch and
return a final state and reward, evaluated in
Section~\ref{subsection:computational-efficiency}. Specifying the observation
and reward spaces as arguments to \texttt{\small step()} enables efficient lazy
computation of observations or rewards in cases where the values are not needed
at every step, or to flexibly change observation and reward space during the
liftetime of an environment.

\subsubsection{Lightweight deep copy operator}

\cg{} environments provide a \texttt{fork()} operator that efficiently creates
independent deep copies of environment states. This can be used to optimize
backtracking or other techniques that require frequently evaluating a common
subsequence of actions. For example, a greedy search can be implemented by
creating $n$ forks of an environment with an $n$-dimensional action space,
running a single action in each fork, and selecting the one which produced the
greatest reward. Backtracking is especially expensive in compilers because most
actions have no ``undo''.

\begin{figure}
  \centering %
  \vspace{-.5em}
  \includegraphics[width=.99\linewidth]{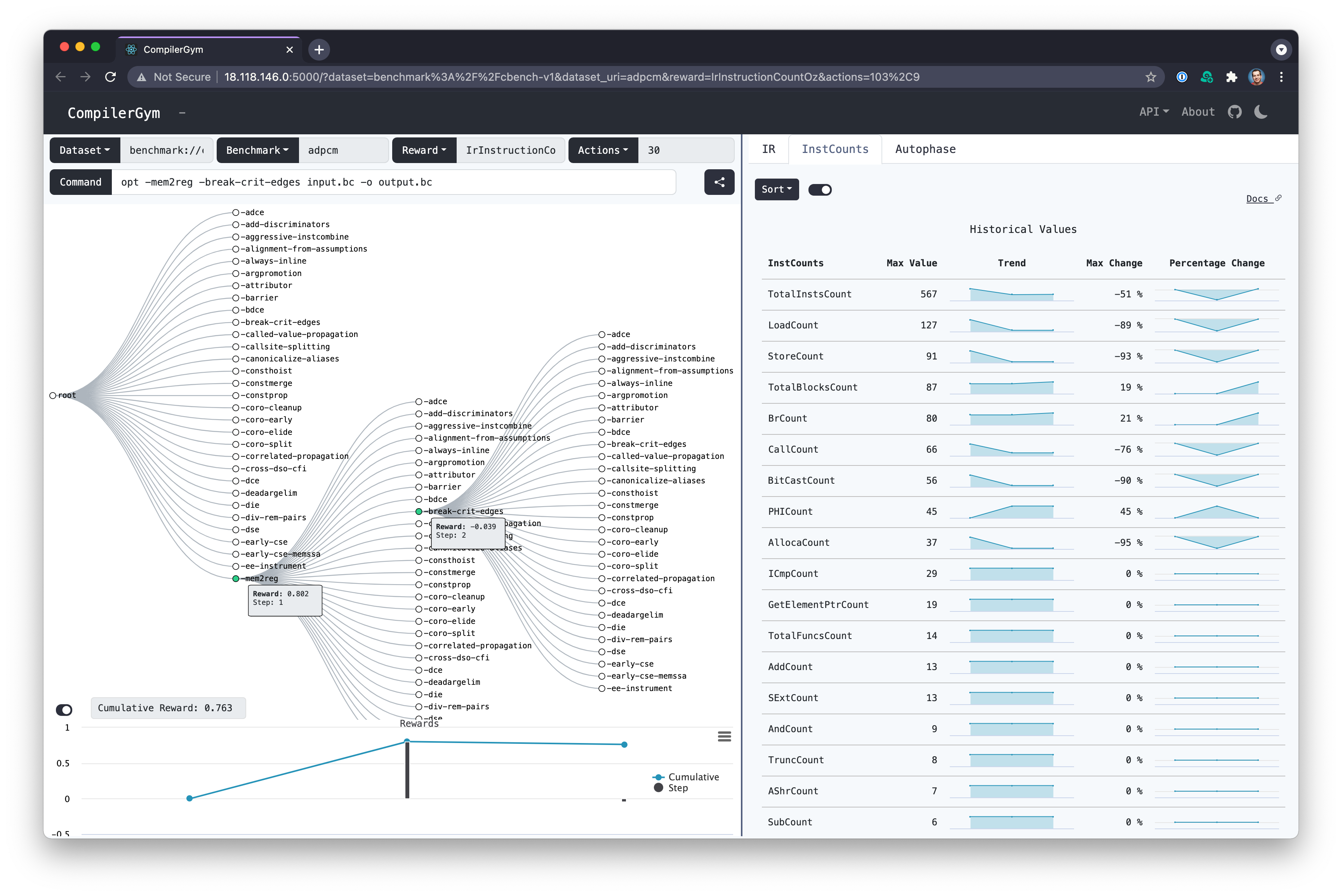}
  \caption{%
    \cg{} Explorer, a website that enables users to interact with the LLVM phase
    ordering environment. The left side of the page renders the phase ordering
    search space as an interactive tree; the right side of the page visualizes
    the program features and their trends.%
    \vspace{-.5em}%
  }%
  \label{figure:web-interface-screenshot}
\end{figure}

\subsection{Customizing Environment Behavior}%
\label{subsection:wrappers}

The Gym~\cite{gym} library defines environment wrapper classes to mutate the MDP
formulation of a wrapped environment. \cg{} provides an additional suite of
environment wrappers for a broad range of compiler research uses. These include
specifying a subset of command line flags to use in an action space, iterating
over a suite of benchmarks, and defining derived observation spaces such as
using custom compiler analyses on compiler IR. These wrappers can be composed.
Listing~\ref{listing:rllib-demo} shows integration with the popular
RLlib~\cite{rllib} library using two of these wrappers.

\subsection{Command Line Tools}

We include a complete set of command line tools for \cg{}, including scripts to
run parallelized searches, replay and validate results from past runs, and an
interactive shell that includes inline documentation and tab completion,
enabling users to interact with the compiler optimization environments without
writing any code.

\subsection{Web Service and \cg{} Explorer}

We designed a REST API to enable \cg{} environments to be used over a network,
and \cg{} Explorer\footnote{Available at:
\url{https://compilergym.ai/explorer}}, a web frontend that makes it easy to
navigate compiler optimization spaces, implemented using React. \cg{} Explorer
presents a visualization of the search tree, shown in
Figure~\ref{figure:web-interface-screenshot}, and asynchronously calls the REST
API to update the tree in real time as the user interacts with it.

A key feature of the tool is to visualize not only the current state, but also
historical trends of the rewards and observation metrics. This allows users to
easily pinpoint interesting actions in a large search tree and trigger new
explorations. We expect this to be valuable for feature engineering, debugging
the behavior of agents, and as a general educational tool.

\subsection{State Transition Dataset}%
\label{subsection:state-transition-dataset}

\begin{figure}
    \centering %
    \includegraphics[width=.7\linewidth]{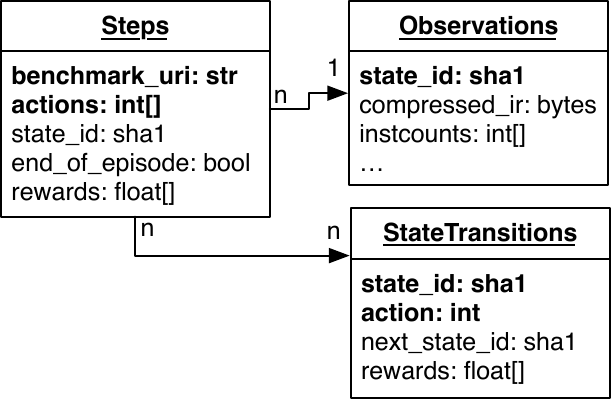}
    \caption{%
        The relational database schema for state transitions in LLVM
        environments. Fields that comprise unique primary keys are emboldened.
        We are releasing an instance of this database comprising over 1M unique
        states that can be used for pre-training, off-policy learning, or
        general offline analysis.%
        \vspace{-.5em}%
    }%
    \label{figure:state-transition-dataset}%
\end{figure}

We designed a relational database schema to log the state transitions of \cg{}
environments for later offline analysis, shown in
Figure~\ref{figure:state-transition-dataset}. A \texttt{\small Steps} table
records every unique action sequence for a particular benchmark and a hash of
the environment state. An \texttt{\small Observations} table stores various
representations of each unique state, indexed by state hash. A \texttt{\small
StateTransitions} table encodes the unique transitions between states and the
rewards received for each.

We implemented a wrapper class for \cg{} environments that asynchronously
populates the \texttt{\small Steps} and \texttt{\small Observations} tables of a
state transition database upon every step of an environment. A post-processing
script de-duplicates and populates the \texttt{\small StateTransitions} table.

We are releasing a large instance of this database ($50+$GB) which contains over
1M unique LLVM environment states, suitable for a range of offline supervised
and unsupervised learning tasks. We evaluate an example usage in
Section~\ref{subsection:sl-offline-dataset}.

\section{Backend Runtime and Interface}%
\label{section:backend}

The \cg{} backend comprises a CompilationSession interface for integrating
compilers and a common client-server runtime that map this interface to the Gym
API.

\subsection{The CompilationSession Interface}

\begin{figure}
  \centering %
  \includegraphics[width=.99\linewidth]{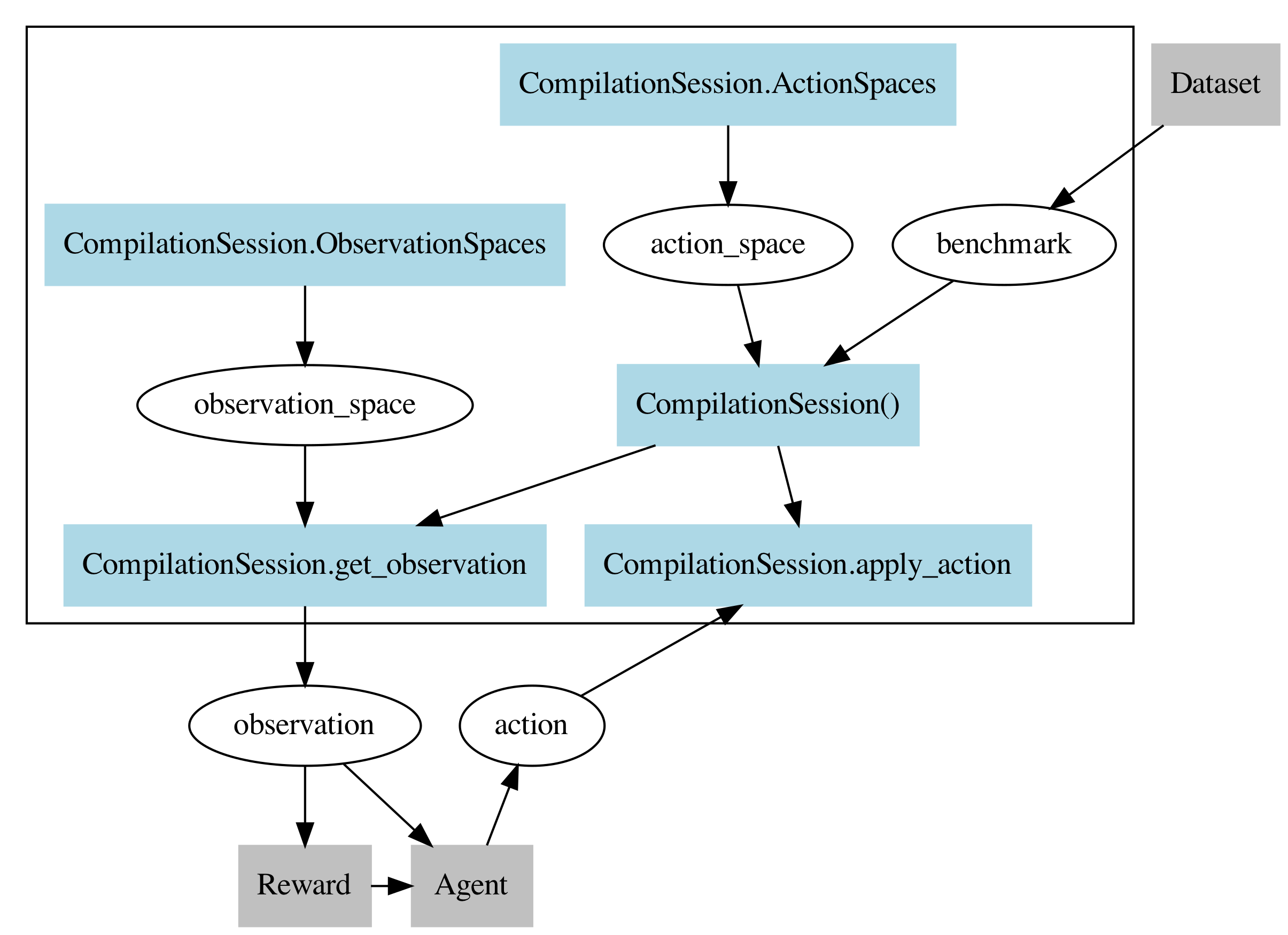}
  \caption{%
    A graphical representation of the CompilationSession integration. To add a
    new compiler to \cg{}, users need only define the boxes highlighted in blue.
    Grey boxes demonstrate the integration of the CompilerSession with a typical
    reinforcement learning loop.%
    \vspace{-.5em}%
  }%
  \label{figure:backend-integration}%
\end{figure}

\cg{} is designed for seamless compiler integration. The integration centers
around implementing a state machine to interact with the compiler called a
CompilationSession. A CompilationSession exposes actions and observations using
a simple schema and must implement two methods, \texttt{\small apply\_action}
and \texttt{\small get\_observation}, as shown in
Figure~\ref{figure:backend-integration}. We provide CompilationSession
interfaces for Python and C++.
Listing~\ref{listing:compilation-session-interface} demonstrates an example
implementation.

\subsection{Compiler Service Runtime}

A common runtime maps implementations of the CompilationSession interface
(Listing~\ref{listing:compilation-session-interface}) to the Gym API
(Listing~\ref{listing:user-api-demo}). This runtime is shared by all compiler
integrations and is architected to be performant and scalable. The design is
resilient to failures, crashes, infinite loops, and nondeterministic behavior in
backend compiler services. All compiler service operations have appropriate
timeouts, graceful error handling, or retry loops. Improvements to the runtime
can be made without changing compiler integration or user code.

A key design point of the \cg{} runtime is that the service that provides the
compiler integration is isolated in a separate process to the user's Python
interpreter. The Python interpreter invokes operations on the compiler service
through Remote Procedure Calls (RPCs). The benefits of this are fault tolerance
and recovery in cases where the compiler crashes or terminates abruptly; support
for compiling on a different system architecture than the host by running the
compiler service on a remote machine; and scalability as the expensive compute
work is offloaded, enabling many user threads to interact with separate compiler
environments without contention on Python's global interpreter lock.

\section{Environments}%
\label{section:environments}

This section describes three compiler integrations shipped in \cg{}.

\subsection{LLVM Phase Ordering}

LLVM~\cite{llvm} is a modular compiler infrastructure used throughout academia
and industry. After parsing an input source program to a language-agnostic
Intermediate Representation (IR), the LLVM optimizer applies a configurable
pipeline of optimization passes to the IR. The selection and ordering of
compiler optimizations -- known as \emph{phase ordering} -- greatly impacts the
quality of the final binary and has been the focus of much
research~\cite{autotuning-survey,evaluating-autotuning}.

We include a phase ordering environment in \cg{} as an example of a challenging,
high-dimensional optimization problem in which significant gains can be
achieved.

\begin{lstfloat}[t]
\begin{lstlisting}[%
    language=C++,%
    caption={%
        A C++ implementation of the CompilationSession interface to add support
        for a new compiler. Method bodies are omitted for brevity. There is an
        equivalent API for Python.%
        \vspace{-.5em}
    },%
    label=listing:compilation-session-interface
]
#include "compiler_gym/service/CompilationSession.h"
#include "compiler_gym/service/runtime/Runtime.h"
using namespace compiler_gym;

struct MyCompilationSession: public CompilationSession{
  vector<ActionSpace> getActionSpaces() {...}
  vector<ObservationSpace> getObservationSpaces() {...}

  Status init(
    const ActionSpace& actionSpace,
    const Benchmark& benchmark) {...}
  Status applyAction(
    const Action& action,
    bool& endOfEpisode,
    bool& actionSpaceChanged) {...}
  Status setObservation(
    const ObservationSpace& observationSpace,
    Observation& observation) {...}
};

int main(int argc, char** argv) {
  runtime::createAndRunService<MyCompilationSession>(
    argc, argv, "My compiler service");
}
\end{lstlisting}
\end{lstfloat}
%
%
%
%
%

\subsubsection*{Actions} The action space consists of a discrete choice from 124
optimization passes extracted automatically from LLVM. There is no maximal
episode length as episodes can run forever (except in the case of a compiler bug
leading to an error), the user must estimate when no further gains can be
achieved and no further actions should be taken. For any particular program the
optimal phase ordering may omit or repeat actions.

\subsubsection*{Rewards}

We support optimizing for three metrics: \emph{code size}, which is the number
of instructions in the IR; \emph{binary size}, which is the size of the
\texttt{.text} section in the compiled object file; and \emph{runtime}, which is
the wall time of the compiled program when run using a specific configuration of
inputs on the machine hosting the \cg{} backend. When used as a reward signal
each metric returns the change in value between the previous environment state
and the new environment state. Each reward signal can optionally be scaled
against the gains achieved by the compiler's default phase orderings,
\texttt{-Oz} for size reduction and \texttt{-O3} for runtime. Code size is
platform-independent and determinsitic, binary size is platform-dependent and
deterministic, and runtime is both platform-specific and nondeterministic.

\subsubsection*{Observations}

We provide five observation spaces for LLVM ranging from counter-based numeric
feature vectors~\cite{autophase} to sequential language models~\cite{inst2vec}
up to graph-based program representations~\cite{programl}. See
Table~\ref{table:llvm-observation-spaces} for a comparison.

\subsubsection*{Datasets}

We provide millions of programs for evaluation, summarized in
Table~\ref{table:llvm-datasets}. We aggregate C, C++, OpenCL, and Fortran
programs from benchmark suites in a variety of domains, open source programs,
and synthetic program generators. Accessing these datasets within \cg{} is as
simple as specifying the name of the dataset and optionally the name of a
specific benchmark. Presently only cBench~\cite{cbench} and Csmith~\cite{csmith}
support optimizing for runtime.

\begin{table}
  \centering
  \footnotesize
  \begin{tabular}{l | r | r | r}
      \toprule
       & \multicolumn{3}{c}{Number of Benchmarks}\\
      Dataset & Autophase~\cite{autophase} & MLGO~\cite{mlgo} & \cg{}\\
      \midrule
      AnghaBench~\cite{anghabench} & & & 1,041,333 \\
      BLAS~\cite{blas} & & & 300\\
      cBench~\cite{cbench} & & & 23 \\
      CHStone~\cite{chstone} & 9 & & 12 \\
      CLgen~\cite{clgen} & & & 996 \\
      GitHub~\cite{programl} & & &  49,738 \\
      Linux kernel & & & 13,894 \\
      MiBench~\cite{mibench} & & &  40 \\
      NPB~\cite{npb} & & &  122 \\
      OpenCV & & &  442 \\
      POJ-104~\cite{poj104} & & &  49,816 \\
      TensorFlow~\cite{tensorflow} & & &  1,985 \\
      Csmith~\cite{csmith} & 100 & & $2^{32}\dagger$ \\
      llvm-stress~\cite{llvm} & & & $2^{32}\dagger$ \\
      Proprietary & & 28,000 & \\
      \bottomrule
  \end{tabular}
  \caption{%
    LLVM benchmark datasets included in \cg{}, compared to the number of
    benchmarks used in two recent machine learning works. $\dagger$ denotes
    random program generators with 32-bit seeds. Excluding the program
    generators, the total number of benchmarks is 1,145,499.%
    \vspace{-.5em}%
  }
  \label{table:llvm-datasets}
\end{table}

\subsection{\gcc Flag Tuning}

We include an environment that exposes the optimization space defined by
\gcc{}'s command line flags. The environment works with any version of \gcc from
5 up to and including the current version at time of writing, 11.2. The
environment uses Docker images to enable hassle free install and consistency
across machines. Alternatively, any local installation of the compiler can be
used. This selection is made by simple string specifier of the path or docker
image name. The only change that an RL agent needs to make to work with \gcc
instead of \llvm is to call \texttt{\small env~=~gym.make("gcc-v0")}, instead of
using \texttt{\small "llvm-v0"}.

While the LLVM phase ordering action space is unbounded as passes may be
executed forever, the number of GCC command line configurations is bounded.
\gcc's action space consists of all the available optimization flags and
parameters that can be specified from the command line. These are automatically
extracted from the ``help'' documentation of whichever \gcc version is used. For
\gcc 11.2.0\footnote{11.2.0 is the latest stable version of \gcc at time of
writing.}, the optimization space includes 502 options:
\begin{itemize}
    \item the six \texttt{\small -O\textit{<n>}} flags, e.g. \texttt{\small -O0,
    -O3, -Ofast, -Os}.
    \item 242 flags such as \texttt{\small -fpeel-loops}, each of which may be
    missing, present, or negated (e.g. \texttt{\small -fno-peel-loops}). Some of
    these flags may take integer or enumerated arguments which are also included
    in the space.
    \item 260 parameterized command line flags such as \texttt{\small -{}-param}
    \texttt{\small inline-heuristics-hint-percent=<number>}.  The number of
    options for each of these varies. Most take numbers, a few take enumerated
    values.
\end{itemize}
This gives a finite optimization space with a modest size of approximately
$10^{4461}$. Earlier versions of \gcc report their parameter spaces less clearly
and so the tool finds smaller spaces when pointed at those. For example, on \gcc
5, the optimization space is only $10^{430}$.

\subsubsection*{Actions}

We provide two action spaces that can be used interchangeably. The first
directly exposes the optimization space via a list of integers, each encoding
the choice for one option with a known cardinality. A second action space is
intended to make it easy for RL tools that operate on a flat list of categorical
actions. For every option with a cardinality of fewer than ten, we provide
actions that directly set the choice for that action. For options with greater
cardinalities we provide actions that add and subtract 1, 10, 100, and 1000 to
the choice integer corresponding to the option. For \gcc 11.2.0, this creates a
set of 2281 actions that can modify the choices of the current state.

\subsubsection*{Rewards}

We provide two deterministic reward signals: the sizes in bytes of the assembly
or the object code.

\subsubsection*{Observations}

We provide four observation spaces: a numeric instruction count observation, the
Register Transfer Language code at the end of compilation, the assembly code as
text, and the object code as a binary.

\subsection{CUDA Loop Nest Code Generation}

Manually tuning CUDA code requires sweeping over many parameters.  Due to the
sheer size of the tunable space, the problem of generating fast CUDA is well
suited for automated techniques~\cite{ansor, halide}. As a flexible compilation
environment, \cg{} is well equipped to handle compilers for tuning GPU
workloads.  We integrated \texttt{loop\_tool}, a simple dense linear algebra
compiler~\cite{loop-tool}. \texttt{loop\_tool} takes a minimalist approach to
linear algebra representations by decomposing standard BLAS-like routines into a
DAG of $n$-dimensional applications of arithmetic primitives.  The DAG is then
annotated with three pieces of information about loop ordering: the order in
which loops are emitted, the nesting structure of each loop, and the reuse of
loops by subsequent operations.  This is lowered to a loop tree that can be
annotated with which loop should be run in parallel. These four annotations
across a slew of point-wise operations represent a large optimization space.

\begin{lstfloat}[t]
\begin{lstlisting}[%
    caption={%
        An example loop tree in the loop\_tool environment.%
        \vspace{-1.5em}%
    },%
    label=listing:loop-tool-manual-session
]
for a in 1048576 : L0 [thread]
 for a' in 1 : L1
  for a'' in 1 : L2
   %0[a] <- read()
  for a'' in 1 : L4
   %1[a] <- read()
  for a'' in 1 : L6
   %2[a] <- add(%0, %1)
  for a'' in 1 : L8
   %3[a] <- write(%2)
  \end{lstlisting}
\end{lstfloat}

\subsubsection*{Actions}

We map interacting with the loop structure for point-wise additions to a
cursor-based discrete action space. At any point the cursor will refer to an
individual loop in the loop hierarchy and will have an associated ``mode'' to
control either moving the cursor or modifying the current loop. There is an
action ``toggle\_mode'' to swap between these two. When moving the cursor, the
actions ``up'' and ``down'' will shift the cursor inward and outward
respectively. When modifying the current loop, the action ``up'' will increase
its size by one. This is done by changing the size of the parent loop to
accommodate the new inner size.  Often this induces tail logic, which is handled
automatically. Finally, any loop can be changed to be threaded.  This will
schedule loop execution across CUDA threads which may span multiple warps or
even multiple streaming multiprocessors. A second, extended action space allows
loops to be split, creating a larger hierarchy.

\subsubsection*{Rewards}

The environment reward signal is a measurement of floating point operations per
second (FLOPs) achieved by benchmarking the loop nest in the given state. This
is both platform dependent and non-deterministic due to the noise involved in
benchmarking.

\subsubsection*{Observations}

There are two observations spaces: action state, which describes the cursor
position and mode, and loop tree structure, which is a textual dump of the
current state of the \texttt{loop\_tool} environment, as shown in
Listing~\ref{listing:loop-tool-manual-session}.
\section{Implementation}

\cg{} is implemented in a mixture of Python and C++. The core runtime comprises
12k lines of code. The compiler integrations comprise 6k lines of code for LLVM,
3k for GCC and 0.5k for \texttt{loop\_tool}. \cg{} is open source and available
under a permissive license.

\subsubsection*{Binary Releases}

Periodic versioned releases are made from a stable branch. We ship pre-compiled
release binaries for macOS and Linux (Ubuntu 18.04, Fedora 28, Debian 10 or
newer equivalents) that can be installed as Python wheels.

\subsubsection*{Documentation}

Our public facing documentation includes full API references for Python and C++,
a getting started guide, FAQ, and code samples demonstrating integration with
RLlib~\cite{rllib}, implementations of exhaustive, random, and greedy searches,
and Q-learning~\cite{q-learning} and Actor Critic~\cite{actor-critic}.

\subsubsection*{Testing}

We have a comprehensive unittest suite with 85.8\% branch coverage that is run
on every code change across a test matrix of all supported operating systems and
Python versions. Additionally, a suite of fuzz and stress tests are ran daily by
continuous integration services to proactively identify issues.

\section{Evaluation}

We evaluate \cg{} first by comparing the computational efficiency of the
environments to prior works. We then show how the simplicity of the \cg{} APIs
enables large-scale autotuning and reinforcement learning experiments to be
engineered with remarkably few lines of code.

\subsubsection*{Experimental Platforms} Results in this section are obtained
from shared compute servers equipped with Intel Xeon 8259CL CPUs, NVIDIA GP100
GPUs, and flash storage.

\subsection{Computational Efficiency}%
\label{subsection:computational-efficiency}

A key design goal of \cg{} is to provide the best performance possible, enabling
researchers to train larger models, try more configurations, and get better
results in less time. We evaluate the computational efficiency of \cg{}'s LLVM
phase ordering environment and compare to two prior works:
Autophase~\cite{autophase} and OpenTuner~\cite{opentuner}.

We use code size rewards signals for all three platforms and the observation
space used in~\cite{autophase} for Autophase and \cg{}; OpenTuner is a black box
search framework and so does not provide observation spaces. We measure the
computational efficiencies of each environment by measuring the wall times of
operations during 1M random trajectories. For \cg{}, which uses a client-server
architecture, we also measure the initial server startup time.

\begin{table*}[t!]
  \centering
  \scriptsize
  \begin{tabular}{l | r r r r | r r r r | r r r r}
      \toprule
      & \multicolumn{4}{c}{Service Startup} & \multicolumn{4}{c}{Environment Initialization} & \multicolumn{4}{c}{Environment Step} \\
      & Cost & p50 & p99 & $\mu$ & Cost & p50 & p99 & $\mu$ & Cost & p50 & p99 & $\mu$\\
      \midrule
      Autophase~\cite{autophase} & \textbf{---} & \textbf{---} & \textbf{---} & \textbf{---} & $\bigo{(n)}$ & 22.4ms & 388.4ms & 53.3ms & $\bigo{(nm)}$ & 71.0ms & 2,489.8ms & 205.9ms\\
      OpenTuner~\cite{opentuner} & \textbf{---} & \textbf{---} & \textbf{---} & \textbf{---} & $\bigo{(n)}$ & 269.6ms & 8,515.3ms & 777.5ms & $\bigo{(nm)}$ & 50.7ms & 1,491.1ms & 131.2ms\\
      \cg{}                      & $\bigo{(1)}$ & 119.7ms & 131.8ms & 120.8ms & $\bm{\bigo{(1)}}\dagger$ & \textbf{2.2ms} & \textbf{198.6ms} & \textbf{21.3ms} & $\bm{\bigo{(n)}}$ & \textbf{1.0ms} & \textbf{108.6ms} & \textbf{7.5ms} \\
      \midrule
      \cg{}-batched & '' & '' & '' & '' & '' & '' & '' & '' & '' & \textbf{0.2ms} & \textbf{37.4ms} & \textbf{2.6ms} \\
      \bottomrule
  \end{tabular}
  \caption{%
      Computational costs of \cg{} operations compared to prior works when
      computing the same actions, observations, and rewards. After paying a
      one-off startup penalty, \cg{} is $27\times$ faster than an equivalent
      prior work. This much higher throughput enables training larger models
      with larger datasets. Cost denotes average-case time complexities
      \textit{wrt}.\ $n$ the size of the program being compiled and $m$ the
      number of actions in the episode. $\dagger$ denotes amortized cost,
      achieved by caching initial environment states. p50 and p99 denote the
      50th and 99th percentile of wall times, respectively. $\mu$ denotes the
      arithmetic mean wall time. We also measured the throughput of \cg{} when
      batches of actions are processed in a single environment step, denoted
      \cg{}-batched above. This improves throughput by a further $2.9\times$ by
      reducing RPC round trips, but loses intermediate observations and rewards.
      Measurements taken from 1M randon trajectories, evenly divided across all
      benchmark datasets.%
      \vspace{-.5em}%
  }
  \label{table:llvm-op-costs}
\end{table*}

Table~\ref{table:llvm-op-costs} shows the results. \cg{} achieves a much higher
throughput than Autophase while offering the same interface, observation space,
and reward signal. This is enabled by \cg{}'s client-server architecture. After
initially reading and parsing the bitcode file from disk, the \cg{} server
incrementally applies an individual optimization pass at each step. In contrast,
Autophase and OpenTuner must, at each step, read and parse the IR, apply the
entire sequence of passes, and then serialize the result. OpenTuner, which was
designed for uses where the search time is dominated by compilation time, has
the highest environment initialization cost, as it requires several disk
operations and the creation of a database. The \cg{} server maintains a cache of
parsed unoptimized bitcodes that enables an amortized $\bigo(1)$ cost of
environment initialization.

\begin{figure}
  \centering %
  \includegraphics[width=.95\linewidth]{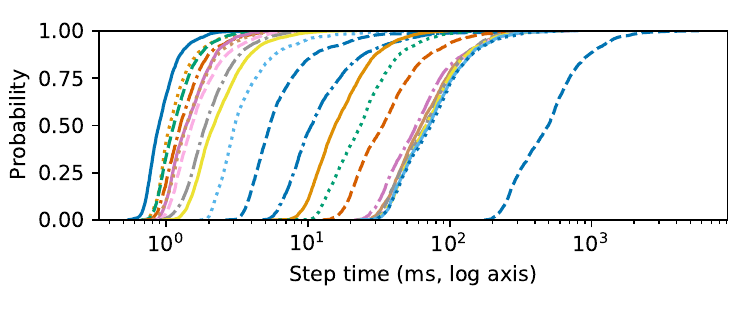}
  \vspace{-1em}
  \caption{%
    Cumulative density plot of step times for each of the 23 programs in
    cBench~\cite{cbench}. Each line shows a different program. The difference
    between the median step times of the fastest program (crc32) and the slowest
    (ghostbench) is 560.3$\times$.%
    \vspace{-.5em}
  }%
  \label{figure:step-time-cdf-cbench}
\end{figure}

The distribution of operation wall times depends on the action being performed
and the program being optimized. Figure~\ref{figure:step-time-cdf-cbench} shows
the wide distribution of wall times within the benchmarks of a single dataset.

\subsection{Computational Efficiency of Observation Spaces}%
\label{subsection:observation-spaces-efficiency}

This experiment evaluates the computational efficiency of the LLVM environment
observation and reward spaces. We recorded 1M wall times of each using random
trajectories.

Table~\ref{table:llvm-observation-spaces} summarizes the results. There is a
$192\times$ range in observation space times, demonstrating a tradeoff between
observation space computational cost and fidelity; and $4727\times$ range in
reward space times, motivating the development of fast approximate proxy rewards
and cost models~\cite{ithemal,halide-cost-model}.

\begin{table}
  \scriptsize
  \begin{tabular}{l l r r r}
    \toprule
    & Type & p50 & p99 & $\mu$\\
    \midrule
    LLVM-IR & String & 0.9ms & 72.1ms & 5.9ms \\
    InstCount & 70-D int64 vector & 0.5ms & 6.9ms & 0.9ms \\
    Autophase~\cite{autophase} & 56-D int64 vector & 0.7ms & 38.0ms & 3.4ms \\
    inst2vec~\cite{inst2vec} & 200-D float vector list & 15.8ms & 31,847ms & 738.1ms\\
    ProGraML~\cite{programl} & Directed multigraph & 104.5ms & 14,194ms & 821.5ms\\
    \midrule
    Code size & Int64 count & 0.4ms & 3.6ms & 0.4ms\\
    Binary size & Int64 byte count & 56.2ms & 703.7ms & 98.1ms\\
    Runtime & Float wall time & 75.9ms & 8,406ms & 614.4ms\\
    \bottomrule
  \end{tabular}
  \caption{%
    Computational costs of the observation and reward spaces of LLVM
    environments from 1M wall time measurements, evenly divided across all
    benchmark datasets. p50 and p99 denote the 50th and 99th percentile,
    respectively, and $\mu$ denotes the arithmetic mean.%
    \vspace{-.5em}%
  }
  \label{table:llvm-observation-spaces}
\end{table}

\subsection{Autotuning LLVM Phase Ordering}%
\label{subsection:autotuning-llvm}

We evaluate various autotuning techniques on the LLVM phase ordering task to
demonstrate the ease and speed of \cg. We use the following autotuning
techniques: Greedy search, which at each step evaluates all possible actions and
selects the action which provides the greatest reward, terminating once no
positive reward can be achieved by any action; LaMCTS~\cite{lamcts}, an
extension of Monte Carlo Tree Search~\cite{mcts} that partitions the search
space on the fly to focus on important search regions~\cite{mcts};
Nevergrad~\cite{nevergrad} and OpenTuner~\cite{opentuner}, two black box
optimization frameworks that contain ensembles of techniques; and Random Search,
which selects actions randomly until a configurable number of steps have elapsed
without a positive reward.

We run single threaded versions of each autotuning technique on each benchmark
in the cBench~\cite{cbench} suite for one hour. Hyperparameters for all
techniques were tuned on a validation set of 50 Csmith~\cite{csmith} benchmarks.
We evaluate each technique when optimizing for three different targets: code
size, binary size, and runtime. For runtime we use the median of three
measurements to provide the reward signals during search, and the median of 30
measurements for final reported values. Each experiment was repeated 10 times.

The standard interface exposed by \cg{} makes it simple to integrate with third
party autotuning libraries or to develop new autotuning approaches.
Table~\ref{table:llvm-autotuning} shows the number of lines of code required to
integrate each search technique, and the performance achieved.

Phase ordering is challenging because the optimization space is unbounded,
high-dimensional, and contains sparse rewards. Nevertheless, autotuning -- when
furnished with a sufficiently generous search budget -- outperforms the default
compiler heuristics by tailoring the configuration to each benchmark. We note
that the optimal configuration differs between all benchmarks and optimization
targets.

\begin{table}
  \footnotesize
  \centering
  \begin{tabular}{l R{.9cm} R{1.2cm} R{1.3cm} R{1.2cm}}
      \toprule
      & Lines of code & Geomean code size reduction & Geomean binary size reduction & Geomean runtime speedup \\
      \midrule
      Greedy Search & 10 & $1.053\times$ & $1.267\times$ & $1.059\times$ \\
      LaMCTS~\cite{lamcts} & 35 & $1.051\times$ & $1.273\times$ & $1.053\times$ \\
      Nevergrad~\cite{nevergrad} & 41 & $\bm{1.083\times}$  & $\bm{1.318\times}$ & $\bm{1.093\times}$ \\
      OpenTuner~\cite{opentuner} & 165 & $1.060\times$ & $1.102\times$ & $0.822\times$ \\
      Random Search & 24 & $1.048\times$ & $1.278\times$ & $1.078\times$ \\
      \bottomrule
  \end{tabular}
  \caption{%
    Lines of code required to integrate or implement autotuning techniques for
    the LLVM phase ordering task, and performance achieved when optimizing for
    three targets on the cBench suite~\cite{cbench} given a 1hr search budget.
    Results are compared against \texttt{-Oz} for size reduction and
    \texttt{-O3} for runtime.%
  }
  \label{table:llvm-autotuning}
\end{table}

\subsection{Autotuning GCC Command Line Flags}%
\label{subsection:autotuning-gcc}

For \gcc we show a different aspect of the \cg. For these experiments we explore
the \gcc environment's high-dimensional action space using a number of simple
search techniques. These experiments are performed using \gcc version 11.2.0 in
Docker. That version of \gcc has 502 optimization settings that can be selected.
We evaluate three search techniques on the the CHstone~\cite{chstone} suite:

1) \emph{Random search.} A random list of 502 integers from the allowable range
is selected at each step.

2) \emph{Hill climbing search.} At each step a small number of random changes
are made to the current choices. If this improves the objective then the current
state is accepted and future steps modify from there.

3) \emph{Genetic algorithm (GA).} A population of 100 random choices is
maintained. We use the Python library \texttt{\small
geneticalgorithm}~\cite{pyga} with its default parameters.

Table~\ref{table:gcc-autotuning} shows the geometric mean of the object code
size objective across the benchmarks in CHstone~\cite{chstone}, averaged over 3
searches. Each search was allowed 1000 compilations.

\subsection{Autotuning CUDA Loop Nests}%
\label{subsection:autotuning-loop-tool}

\begin{figure}
  \centering %
  \vspace{-.5em}
  \includegraphics[width=.95\linewidth]{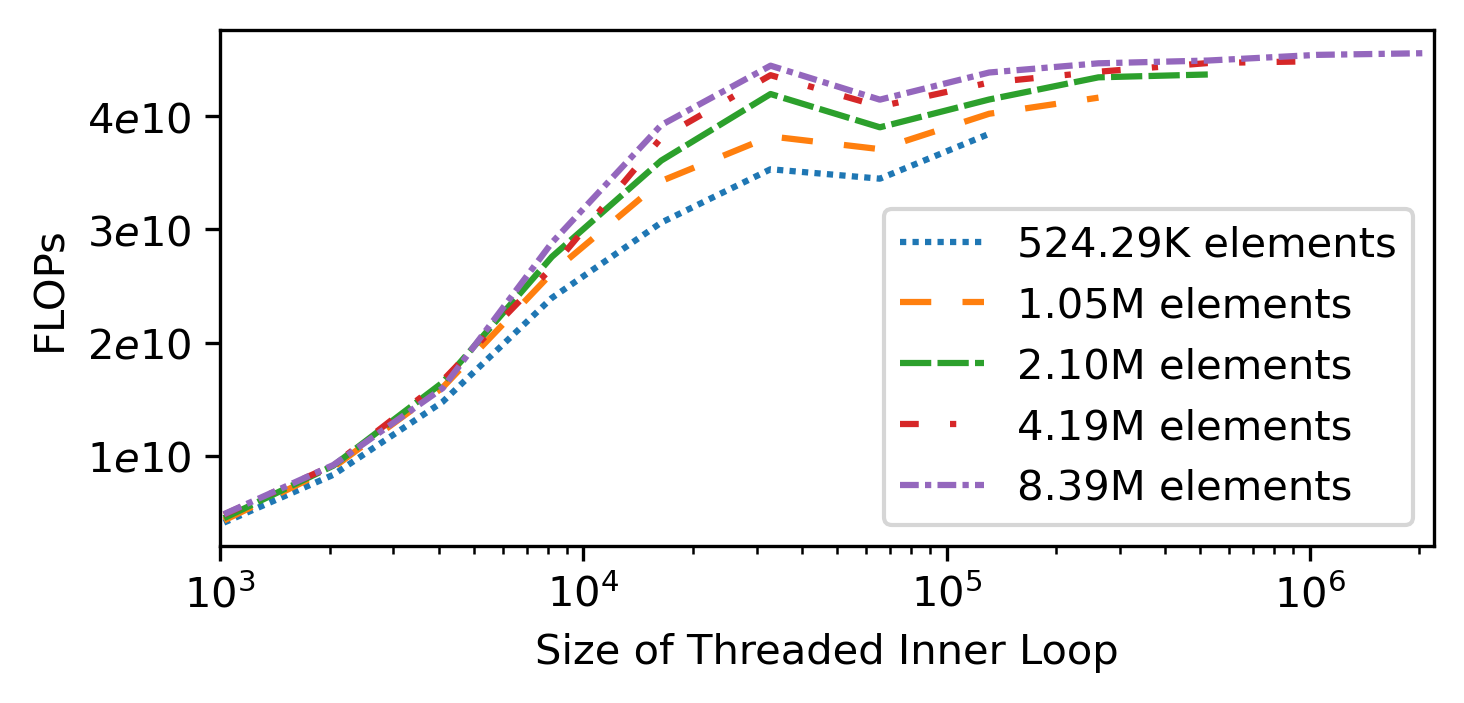}
  \vspace{-1em}
  \caption{%
      A sweep over a particular configuration in the \cg{} provided search space
      for point-wise addition on a GPU using \texttt{\small loop\_tool} to
      generate CUDA.%
  }%
  \label{figure:loop-tool-sweep}
\end{figure}

The \texttt{loop\_tool} environment provides an easily accessible interface to
being exploring the landscape of GPU optimizations. Tuning a simple space by
searching threading and then sizing the inner loop reaches 73.5\% of theoretical
peak performance on our GP100 test hardware ($\sim$6e10 FLOPs or $\sim$750GB/s
for two 4-byte floating point reads and one write), and parity with PyTorch
performance on the same operation across a variety of problem sizes.
Figure~\ref{figure:loop-tool-sweep} shows the results for different loop
configurations, demonstrating potentially useful hardware and compiler
characteristics, notably a drop in performance near 100k threads.

\begin{table}
  \footnotesize
  \centering
  \begin{tabular}{l r R{3cm}}
      \toprule
      & Lines of code & Geomean binary size reduction\\
      \midrule
      Genetic Algorithm~\cite{pyga} & 27 & $\bm{1.27\times}$ \\
      Hill Climbing & 14 & $1.04\times$ \\
      Random Search & 9 & $1.21\times$ \\
      \bottomrule
  \end{tabular}
  \caption{%
        Lines of code required to integrate or implement autotuning techniques
        for the GCC flag tuning task, and performance achieved when optimizing
        the CHstone suite~\cite{chstone}, given a search budget of 1000
        compilations per benchmark. Results are compared against \texttt{-Os}.%
        \vspace{-.2em}%
  }
  \label{table:gcc-autotuning}
\end{table}

\subsection{Learning a Cost Model using the State Transition Dataset}%
\label{subsection:sl-offline-dataset}

\begin{figure}[t]
  \centering %
  \vspace{-.8em}
  \includegraphics[width=.9\linewidth]{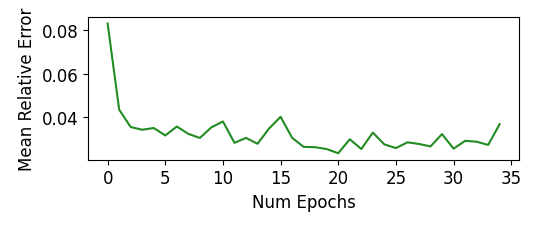}
  \vspace{-.8em}
  \caption{%
    Predicting program instruction using a Graph Neural Networks trained on
    \cg{}'s State Transition Dataset. This shows performance on a holdout
    validation set as a function of the number of training epochs.
  }%
  \label{figure:supervised-learning}%
\end{figure}

Auxiliary tasks are commonly used in reinforcement learning to produce better
representation learning and help with downstream tasks
\cite{Optimal-representations-auxiliary,RL-unsupervised-aux}. This experiment
demonstrates using the State Transition Dataset
(Section~\ref{subsection:state-transition-dataset}) to learn a cost model of
instruction count from a graph representation of program state.

We implemented a Gated Graph Neural Network~\cite{ggnn} in PyTorch~
\cite{pytorch} and used Mean Squared Error loss to train a regressor to predict
the instruction count of a program after two rounds of message passing on the
\pg{}~\cite{programl} graph representation built into \cg{}. We trained on 80\%
of the State Transition Database by iterating over pairs of (graph, instruction
count) from the database. We used the remaining 20\% of the database as a
validation set. Figure~\ref{figure:supervised-learning} shows the convergence of
the neural network. The network achieves a relative error of 0.025, while a
naive mean prediction scores 1.393.

\subsection{Reinforcement Learning for LLVM Phase Ordering}%
\label{subsection:rl-algos}

\begin{table}
    \centering
    \scriptsize
    \begin{tabular}{l | r r r r}
        \toprule
        & \multicolumn{4}{c}{Geomean code size reduction} \\
        Test Dataset & A2C~\cite{a2c} & APEX~\cite{apex} & IMPALA~\cite{impala} & PPO~\cite{ppo} \\
        \midrule
        AnghaBench~\cite{anghabench} & $0.951\times$ & $0.659\times$ & $\bm{0.958\times}$ & $0.776\times$ \\
        BLAS~\cite{blas} & $0.928\times$ & $\bm{0.934\times}$ & $0.861\times$ & $0.906\times$\\
        cBench~\cite{cbench} & $0.804\times$ & $0.698\times$ & $0.814\times$ & $\bm{0.964\times}$ \\
        CHStone~\cite{chstone} & $0.823\times$ & $0.704\times$ & $0.707\times$ & $\bm{1.014\times}$ \\
        CLgen~\cite{clgen} & $\bm{0.950\times}$ & $0.687\times$ & $0.916\times$ & $0.843\times$ \\
        Csmith~\cite{csmith} & $1.023\times$ & $0.692\times$ & $1.144\times$ & $\bm{1.245\times}$ \\
        GitHub~\cite{programl} & $0.975\times$ & $\bm{0.987\times}$ & $0.976\times$ & $0.984\times$ \\
        Linux kernel & $0.987\times$ & $\bm{0.998\times}$ & $0.983\times$ & $0.995\times$ \\
        llvm-stress~\cite{llvm} & $\bm{0.838\times}$ & $0.493\times$ & $0.736\times$ & $0.097\times$ \\
        MiBench~\cite{mibench} & $0.996\times$ & $0.996\times$ & $0.996\times$ & $\bm{1.000\times}$ \\
        NPB~\cite{npb} & $\bm{0.961\times}$ & $0.816\times$ & $0.958\times$ & $0.923\times$ \\
        OpenCV & $0.976\times$ & $0.969\times$ & $\bm{0.986\times}$ & $0.945\times$ \\
        POJ-104~\cite{poj104} & $0.778\times$ & $0.651\times$ & $\bm{0.805\times}$ & $0.801\times$ \\
        TensorFlow~\cite{tensorflow} & $0.976\times$ & $\bm{0.976\times}$ & $0.966\times$ & $0.933\times$ \\
        \bottomrule
    \end{tabular}
    \caption{%
        Comparison of four reinforcement learning algorithms, trained for 100k
        episodes on Csmith~\cite{csmith} programs, when evaluated on datasets
        from a range of program domains. The programs used for testing on Csmith
        are different from those used for training. Results are compared to
        \texttt{-Oz}. }
    \label{table:rl-algos}
\end{table}

\cg{} offers seamless integration with third party reinforcement learning
frameworks. For example, by changing a single parameter value in
Listing~\ref{listing:rllib-demo} we can use any of the 26 reinforcement learning
algorithms included in RLlib~\cite{rllib}.

We use \cg{} to replicate the LLVM phase ordering environment used
in~\cite{autophase}. Specifically: we fix episode lengths to 45 steps, use the
same observation space comprising a feature vector concatenated with a histogram
of the agent's previous actions, and we use a subset of the full action
space\footnote{We use 42 actions (out of 124 total) rather than the 45 actions
used in~\cite{autophase} as three of the actions have been removed in recent
versions of LLVM.}. We note that each of these modifications to the base LLVM
environment can be achieved using the wrapper classes built into \cg{}
(Section~\ref{subsection:wrappers}). Our environment differs
from~\cite{autophase} in that we use a code size reward signal rather than
simulated cycle counts.

We train three different reinforcement learning algorithms for 100k episodes and
periodically evaluate performance on a holdout validation set. We use Csmith to
generate both training and validation sets, as in~\cite{autophase}.

Table~\ref{table:rl-algos} shows the performance of the trained agents when
evaluated on a random 50 programs from each of the datasets available out of the
box in \cg{}. 3 of the 4 algorithms achieve positive results when generalizing
to programs within the same domain (Csmith), but only PPO~\cite{ppo} is able to
achieve a positive score on two of the 13 other datasets. This highlights the
challenge of generalization across program domains.

\subsection{Effect of Training Set on RL}%
\label{subsection:rl-training-set}

The generalization of reinforcement learning agents across domains is the
subject of active
research~\cite{rl-generalization,rl-generalization-regularization,model-uncertainty}.
As demonstrated in the previous experiment, the performance of agents trained on
one dataset can differ wildly on datasets from other domains. We evaluate the
effect of training set on generalization by training a PPO~\cite{ppo} agent on
different training sets and then evaluating their generalization performance on
test sets from different domains. All other experimental parameters are as per
the previous experiment.

Table~\ref{table:rl-training-set} shows the results. As can be seen, each
algorithm performs best when generalizing to benchmarks from within the same
dataset, suggesting the importance of training on benchmarks across a wide range
of program domains.

\begin{table}
    \footnotesize
    \centering
    \begin{tabular}{r r | R{1.35cm} R{1.35cm} R{2cm}}
        \toprule
        & & \multicolumn{3}{c}{Training Set}\\
        & & Csmith~\cite{csmith} & Github~\cite{programl} & TensorFlow~\cite{tensorflow} \\
        \midrule
        \parbox[t]{2mm}{\multirow{3}{*}{\rotatebox[origin=c]{90}{Test Set}}} & Csmith~\cite{csmith} & $\bm{1.245\times}$ & $0.567\times$ & $0.723\times$ \\
        & Github~\cite{programl} & $0.984\times$ & $\bm{0.981\times}$ & $0.995\times$ \\
        & TensorFlow~\cite{tensorflow} & $0.932\times$ & $0.950\times$ & $\bm{0.998\times}$ \\
        \bottomrule
    \end{tabular}
    \caption{%
        \cg{} includes millions of programs to train on that can be selected by
        simply specifying the name of the dataset(s) to use. Here we
        cross-validate the generalization performance of a PPO~\cite{ppo} agent
        by varying the training and test sets. The row indicates the dataset
        used for training, the column indicates the dataset used for testing.
        The values are geomean code size reduction relative to \texttt{-Oz}.%
    }
    \label{table:rl-training-set}
  \end{table}

\subsection{Effect of Program Representation on Learning}%
\label{subsection:rl-observation-spaces}

Representation learning and feature engineering is an area of much
research~\cite{contextual-embeddings,code2vec-generalizability,ml4sysreview}.
\cg{} environments provide multiple state representations for each environment.
We evaluate the performance of two different program representations, and their
performance when concatenated with a histogram of the agent's previous actions,
as used in~\cite{autophase}. We use the same experimental setup as in the prior
sections.

\begin{figure}
    \centering %
    \includegraphics[width=.98\linewidth]{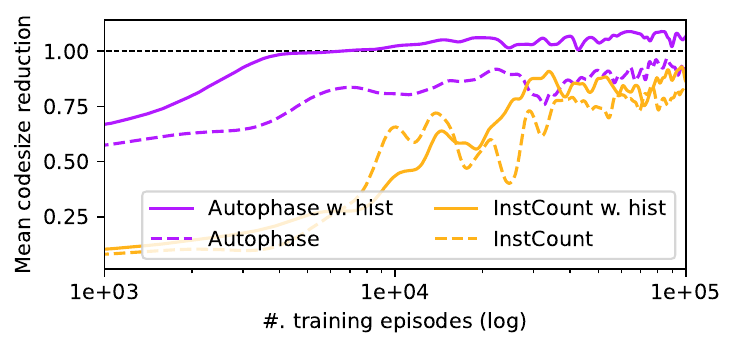}
    \vspace{-.5em}
    \caption{%
        The convergence rate and final performance of learned agents depends on
        the observation space. \cg{} includes several observation spaces out of
        the box. By changing one line of code we trained four PPO~\cite{ppo}
        agents on different observation spaces. This plot shows the performance
        on a holdout validation set as a function of the number of training
        episodes. We applied a Gaussian filter ($\sigma=5$) to aid in
        visualizing the trends. \vspace{-.5em}%
    }
    \label{figure:rl-observation-spaces}
\end{figure}

The results are shown in Figure~\ref{figure:rl-observation-spaces}. In both
cases stronger performance is achieved when coupling the program representation
with a histogram of the agent's previous actions. The Autophase representation
encodes more attributes of the structure of programs than InstCount and achieves
greater performance. We believe that representation learning is one of the most
exciting areas for future research, and \cg{} provides the supporting
infrastructure for this research.

\section{Related Work}

We present a suite of tools for compiler optimization research. Other compiler
research tools include OpenTuner~\cite{opentuner} and YaCoS~\cite{yacos},
autotuning frameworks that include an ensemble of techniques for compiler
optimizations; cTuning~\cite{ctuning}, a framework for distributing autotuning
results; TenSet~\cite{tenset} and LS-CAT~\cite{lscat-dataset}, large-scale
performance datasets suitable for offline learning; and
ComPy-Learn~\cite{compylearn}, a library of program representations for LLVM.
\cg{} has a broader set of features than these prior works, providing several
compiler problems, program representations, optimization targets, and offline
datasets all in a single package.

There is a growing body of research that applies AI techniques to compilers
optimizations~\cite{ml4sysreview}. Many approaches have been proposed to phase
ordering, including collaborative
filtering~\cite{collaborative-filtering-phase-ordering}, design space
exploration~\cite{graph-based-phase-ordering}, and Bayesian
Networks~\cite{cobayn}. Even removing passes from standard optimization
pipelines has been shown to sometimes improve performance~\cite{less-is-more}.
Autophase~\cite{autophase} and CORL~\cite{corl} use reinforcement learning to
tackle the LLVM phase ordering problem. Both works identify generalization
across programs as a key challenge. Our work aims to accelerate progress on this
problem by combining several observation spaces with millions of training
programs to serve as a platform for research.

Other reinforcement learning compiler works include MLGO~\cite{mlgo} which
learns a policy for a function inling heuristic,
NeuroVectorizer~\cite{neurovectorizer} which formulates instruction
vectorization as single-step environments, and PolyGym~\cite{polygym} which
targets Polyhedral loop transformations. Compared to these works, the search
spaces in \cg{} environments are far larger.

\cg{} is not limited to reinforcement learning. Prior work has cast compiler
optimization tasks as supervised learning problems using classification to
select optimization decisions~\cite{deeptune,ml4sysreview} or regression to
learn cost models~\cite{loop-learner,ithemal,halide-cost-model}. \cg{} is as an
ideal platform for gathering the data to train and evaluate these approaches,
including both offline datasets and the infrastructure to generate new ones.

\section{Conclusions}

We aim to lower the barrier-to-entry to compiler optimization research. We
present \cg{}, a suite of tools that removes the significant engineering
investment required try out new ideas on production compiler problems.


\bibliographystyle{unsrt}
\bibliography{references}

\begin{thebibliography}{10}

\bibitem{autotuning-survey}
Amir~H Ashouri, William Killian, John Cavazos, Gianluca Palermo, and Cristina
  Silvano.
\newblock {A Survey on Compiler Autotuning using Machine Learning}.
\newblock {\em CSUR}, 51(5), 2018.

\bibitem{ml4sysreview}
Hugh Leather and Chris Cummins.
\newblock {Machine Learning in Compilers: Past, Present and Future}.
\newblock In {\em FDL}, 2020.

\bibitem{mlgo}
Mircea Trofin, Yundi Qian, Eugene Brevdo, Zinan Lin, Krzysztof Choromanski, and
  David Li.
\newblock {MLGO: a Machine Learning Guided Compiler Optimizations Framework}.
\newblock {\em arXiv:2101.04808}, 2021.

\bibitem{autophase}
Ameer Haj-Ali, Qijing Huang, William Moses, John Xiang, John Wawrzynek, Krste
  Asanovic, and Ion Stoica.
\newblock Autophase: Juggling hls phase orderings in random forests with deep
  reinforcement learning.
\newblock In {\em MLSys}, 2020.

\bibitem{neurovectorizer}
Ameer Haj-Ali, Nesreen~K Ahmed, Ted Willke, Yakun~Sophia Shao, Krste Asanovic,
  and Ion Stoica.
\newblock Neurovectorizer: End-to-end vectorization with deep reinforcement
  learning.
\newblock In {\em CGO}, 2020.

\bibitem{big-code-survey}
Miltiadis Allamanis, Earl~T Barr, Premkumar Devanbu, and Charles Sutton.
\newblock {A Survey of Machine Learning for Big Code and Naturalness}.
\newblock {\em CSUR}, 51(4), 2018.

\bibitem{gym}
Greg Brockman, Vicki Cheung, Ludwig Pettersson, Jonas Schneider, John Schulman,
  Jie Tang, and Wojciech Zaremba.
\newblock {OpenAI Gym}.
\newblock {\em arXiv:1606.01540}, 2016.

\bibitem{csmith}
Xuejun Yang, Yang Chen, Eric Eide, and John Regehr.
\newblock {Finding and Understanding Bugs in C Compilers}.
\newblock In {\em PLDI}, 2011.

\bibitem{rllib}
Eric Liang, Richard Liaw, Robert Nishihara, Philipp Moritz, Roy Fox, Ken
  Goldberg, Joseph Gonzalez, Michael Jordan, and Ion Stoica.
\newblock {RLlib: Abstractions for Distributed Reinforcement Learning}.
\newblock In {\em ICML}, 2018.

\bibitem{difftesting}
William~M McKeeman.
\newblock {Differential Testing for Software}.
\newblock {\em Digital Technical Journal}, 10(1), 1998.

\bibitem{llvm}
Chris Lattner and Vikram Adve.
\newblock {LLVM: A Compilation Framework for Lifelong Program Analysis \&
  Transformation}.
\newblock In {\em CGO}, 2004.

\bibitem{evaluating-autotuning}
Yang Chen, Yuanjie Huang, Lieven Eeckhout, Grigori Fursin, Liang Peng, Olivier
  Temam, and Chengyong Wu.
\newblock {Evaluating Iterative Optimization Across 1000 Datasets}.
\newblock In {\em PLDI}, 2010.

\bibitem{inst2vec}
Tal Ben-Nun, Alice~Shoshana Jakobovits, and Torsten Hoefler.
\newblock {Neural Code Comprehension: A Learnable Representation of Code
  Semantics}.
\newblock In {\em NeurIPS}, 2018.

\bibitem{programl}
Chris Cummins, Zacharias Fisches, Tal Ben-Nun, Torsten Hoefler, Michael
  O'Boyle, and Hugh Leather.
\newblock {ProGraML: A Graph-based Program Representation for Data Flow
  Analysis and Compiler Optimizations}.
\newblock In {\em ICML}, 2021.

\bibitem{cbench}
Grigori Fursin, John Cavazos, Michael O’Boyle, and Olivier Temam.
\newblock {MiDataSets: Creating the conditions for a more realistic evaluation
  of iterative optimization}.
\newblock In {\em HiPEAC}, 2007.

\bibitem{anghabench}
Anderson~Faustino da~Silva, Bruno~Conde Kind, Jos{\'e}~Wesley
  de~Souza~Magalh{\~a}es, Jer{\^o}nimo~Nunes Rocha, Breno Campos~Ferreira
  Guimaraes, and Fernando Magno~Quin{\~a}o Pereira.
\newblock {AnghaBench: A Suite with One Million Compilable C Benchmarks for
  Code-Size Reduction}.
\newblock In {\em CGO}, 2021.

\bibitem{blas}
Chuck~L Lawson, Richard~J. Hanson, David~R Kincaid, and Fred~T. Krogh.
\newblock {Basic Linear Algebra Subprograms for Fortran Usage}.
\newblock {\em TOMS}, 5(3), 1979.

\bibitem{chstone}
Yuko Hara, Hiroyuki Tomiyama, Shinya Honda, Hiroaki Takada, and Katsuya Ishii.
\newblock {CHStone: A Benchmark Program Suite for Practical C-based High-Level
  Synthesis}.
\newblock In {\em ISCAS}, 2008.

\bibitem{clgen}
Chris Cummins, Pavlos Petoumenos, Zheng Wang, and Hugh Leather.
\newblock {Synthesizing Benchmarks for Predictive Modeling}.
\newblock In {\em CGO}, 2017.

\bibitem{mibench}
Matthew~R Guthaus, Jeffrey~S Ringenberg, Dan Ernst, Todd~M Austin, Trevor
  Mudge, and Richard~B Brown.
\newblock {MiBench: A Free, Commercially Representative Embedded Benchmark
  Suite}.
\newblock In {\em WWC}, 2001.

\bibitem{npb}
David Bailey, Tim Harris, William Saphir, Rob Van Der~Wijngaart, Alex Woo, and
  Maurice Yarrow.
\newblock {The NAS Parallel Benchmarks 2.0}.
\newblock Technical report, Technical Report NAS-95-020, NASA Ames Research
  Center, 1995.

\bibitem{poj104}
Lili Mou, Ge~Li, Lu~Zhang, Tao Wang, and Zhi Jin.
\newblock {Convolutional Neural Networks Over Tree Structures for Programming
  Language Processing}.
\newblock In {\em AAAI}, 2016.

\bibitem{tensorflow}
Mart{\'\i}n Abadi, Paul Barham, Jianmin Chen, Zhifeng Chen, Andy Davis, Jeffrey
  Dean, Matthieu Devin, Sanjay Ghemawat, Geoffrey Irving, Michael Isard, et~al.
\newblock {TensorFlow: A System for Large-Scale Machine Learning}.
\newblock In {\em OSDI}, 2016.

\bibitem{ansor}
Lianmin Zheng, Chengfan Jia, Minmin Sun, Zhao Wu, Cody~Hao Yu, Ameer Haj-Ali,
  Yida Wang, Jun Yang, Danyang Zhuo, Koushik Sen, et~al.
\newblock {Ansor: Generating High-Performance Tensor Programs for Deep
  Learning}.
\newblock In {\em OSDI}, 2020.

\bibitem{halide}
Jonathan Ragan-Kelley, Connelly Barnes, Andrew Adams, Sylvain Paris, Fr{\'e}do
  Durand, and Saman Amarasinghe.
\newblock {Halide: A Language and Compiler for Optimizing Parallelism,
  Locality, and Recomputation in Image Processing Pipelines}.
\newblock In {\em PLDI}, 2013.

\bibitem{loop-tool}
Bram Wasti.
\newblock loop\_tool.
\newblock \url{https://github.com/facebookresearch/loop_tool}, 2021.

\bibitem{q-learning}
Christopher~JCH Watkins and Peter Dayan.
\newblock {Q-Learning}.
\newblock {\em Machine learning}, 8(3-4), 1992.

\bibitem{actor-critic}
Vijay~R Konda and John~N Tsitsiklis.
\newblock {Actor-Critic Algorithms}.
\newblock In {\em NeurIPS}, 2000.

\bibitem{opentuner}
Jason Ansel, Shoaib Kamil, Kalyan Veeramachaneni, Jonathan Ragan-Kelley,
  Jeffrey Bosboom, Una-May O'Reilly, and Saman Amarasinghe.
\newblock {OpenTuner: An Extensible Framework for Program Autotuning}.
\newblock In {\em PACT}, 2014.

\bibitem{ithemal}
Charith Mendis, Alex Renda, Saman Amarasinghe, and Michael Carbin.
\newblock {Ithemal: Accurate, Portable and Fast Basic Block Throughput
  Estimation using Deep Neural Networks}.
\newblock In {\em ICML}, 2019.

\bibitem{halide-cost-model}
Benoit Steiner, Chris Cummins, Horace He, and Hugh Leather.
\newblock {Value Learning for Throughput Optimization of Deep Learning
  Workloads}.
\newblock In {\em MLSys}, 2021.

\bibitem{lamcts}
Linnan Wang, Rodrigo Fonseca, and Yuandong Tian.
\newblock {Learning Search Space Partition for Black-Box Optimization using
  Monte Carlo Tree Search}.
\newblock In {\em NeurIPS}, 2020.

\bibitem{mcts}
Guillaume Chaslot, Sander Bakkes, Istvan Szita, and Pieter Spronck.
\newblock {Monte-Carlo Tree Search: A New Framework for Game AI}.
\newblock {\em AIIDE}, 8, 2008.

\bibitem{nevergrad}
Jeremy Rapin and Olivier Teytaud.
\newblock {Nevergrad - A Gradient-Free Optimization Platform}.
\newblock \url{https://github.com/facebookresearch/nevergrad}, 2018.

\bibitem{pyga}
{Ryan Solgi}.
\newblock geneticalgorithm.
\newblock \url{https://pypi.org/project/geneticalgorithm/}, 2020.

\bibitem{Optimal-representations-auxiliary}
Marc~G. Bellemare, Will Dabney, Robert Dadashi, Adrien~Ali Ta{\"{\i}}ga,
  Pablo~Samuel Castro, Nicolas~Le Roux, Dale Schuurmans, Tor Lattimore, and
  Clare Lyle.
\newblock {A Geometric Perspective on Optimal Representations for Reinforcement
  Learning}.
\newblock {\em CoRR}, abs/1901.11530, 2019.

\bibitem{RL-unsupervised-aux}
Max Jaderberg, Volodymyr Mnih, Wojciech~Marian Czarnecki, Tom Schaul, Joel~Z.
  Leibo, David Silver, and Koray Kavukcuoglu.
\newblock {Reinforcement Learning with Unsupervised Auxiliary Tasks}.
\newblock {\em CoRR}, abs/1611.05397, 2016.

\bibitem{ggnn}
Yujia Li, Daniel Tarlow, Marc Brockschmidt, and Richard Zemel.
\newblock {Gated Graph Sequence Neural Networks}.
\newblock {\em arXiv:1511.05493}, 2015.

\bibitem{pytorch}
Adam Paszke, Sam Gross, Francisco Massa, Adam Lerer, James Bradbury, Gregory
  Chanan, Trevor Killeen, Zeming Lin, Natalia Gimelshein, Luca Antiga, et~al.
\newblock {PyTorch: An Imperative Style, High-performance Deep Learning
  Library}.
\newblock {\em arXiv:1912.01703}, 2019.

\bibitem{a2c}
Volodymyr Mnih, Adria~Puigdomenech Badia, Mehdi Mirza, Alex Graves, Timothy
  Lillicrap, Tim Harley, David Silver, and Koray Kavukcuoglu.
\newblock {Asynchronous Methods for Deep Reinforcement Learning}.
\newblock In {\em ICML}, 2016.

\bibitem{apex}
Dan Horgan, John Quan, David Budden, Gabriel Barth-Maron, Matteo Hessel, Hado
  Van~Hasselt, and David Silver.
\newblock {Distributed Prioritized Experience Replay}.
\newblock In {\em ICML}, 2018.

\bibitem{impala}
Lasse Espeholt, Hubert Soyer, Remi Munos, Karen Simonyan, Vlad Mnih, Tom Ward,
  Yotam Doron, Vlad Firoiu, Tim Harley, Iain Dunning, et~al.
\newblock {IMPALA: Scalable Distributed Deep-RL with Importance Weighted
  Actor-Learner Architectures}.
\newblock In {\em ICML}, 2018.

\bibitem{ppo}
John Schulman, Filip Wolski, Prafulla Dhariwal, Alec Radford, and Oleg Klimov.
\newblock {Proximal Policy Optimization Algorithms}.
\newblock {\em arXiv:1707.06347}, 2017.

\bibitem{rl-generalization}
Karl Cobbe, Oleg Klimov, Chris Hesse, Taehoon Kim, and John Schulman.
\newblock {Quantifying Generalization in Reinforcement Learning}.
\newblock In {\em ICML}, 2019.

\bibitem{rl-generalization-regularization}
Kaixin Wang, Bingyi Kang, Jie Shao, and Jiashi Feng.
\newblock {Improving Generalization in Reinforcement Learning with Mixture
  Regularization}.
\newblock In {\em NeurIPS}, 2020.

\bibitem{model-uncertainty}
Yaniv Ovadia, Emily Fertig, Jie Ren, Zachary Nado, David Sculley, Sebastian
  Nowozin, Joshua~V Dillon, Balaji Lakshminarayanan, and Jasper Snoek.
\newblock {Can You Trust Your Model's Uncertainty? Evaluating Predictive
  Uncertainty under Dataset Shift}.
\newblock In {\em NeurIPS}, 2019.

\bibitem{contextual-embeddings}
Aditya Kanade, Petros Maniatis, Gogul Balakrishnan, and Kensen Shi.
\newblock {Learning and Evaluating Contextual Embedding of Source Code}.
\newblock In {\em ICML}, 2020.

\bibitem{code2vec-generalizability}
Hong~Jin Kang, Tegawend{\'e}~F Bissyand{\'e}, and David Lo.
\newblock {Assessing the Generalizability of code2vec Token Embeddings}.
\newblock In {\em ASE}, 2019.

\bibitem{yacos}
Andr{\'e}~Felipe Zanella, Anderson~Faustino da~Silva, and Fernando~Magno
  Quint{\~a}o.
\newblock {YaCoS: a Complete Infrastructure to the Design and Exploration of
  Code Optimization Sequences}.
\newblock In {\em SBLP}, 2019.

\bibitem{ctuning}
Grigori Fursin.
\newblock {Collective Tuning Initiative: Automating and Accelerating
  Development and Optimization of Computing Systems}.
\newblock In {\em GCC Developers' Summit}, 2009.

\bibitem{tenset}
Lianmin Zheng, Ruochen Liu, Ameer~Haj Ali, Junru Shao, Tianqi Chen, Joseph~E
  Gonzalez, and Ion Stoica.
\newblock {TenSet: A Large-scale Program Performance Dataset for Learned Tensor
  Compilers}.
\newblock In {\em NeurIPS}, 2021.

\bibitem{lscat-dataset}
Lars Bjertnes, Jacob~O T{\o}rring, and Anne~C Elster.
\newblock {LS-CAT: A Large-Scale CUDA AutoTuning Dataset}.
\newblock {\em arXiv:2103.14409}, 2021.

\bibitem{compylearn}
Alexander Brauckmann, Andr{\'e}s Goens, and Jeronimo Castrillon.
\newblock {ComPy-Learn}: A toolbox for exploring machine learning
  representations for compilers.
\newblock In {\em FDL}, 2020.

\bibitem{collaborative-filtering-phase-ordering}
Stefano Cereda, Gianluca Palermo, Paolo Cremonesi, and Stefano Doni.
\newblock {A Collaborative Filtering Approach for the Automatic Tuning of
  Compiler Optimisations}.
\newblock In {\em LCTES}, 2020.

\bibitem{graph-based-phase-ordering}
Ricardo Nobre, Luiz~GA Martins, and Jo{\~a}o~MP Cardoso.
\newblock A graph-based iterative compiler pass selection and phase ordering
  approach.
\newblock In {\em LCTES}, 2016.

\bibitem{cobayn}
Amir~Hossein Ashouri, Giovanni Mariani, Gianluca Palermo, Eunjung Park, John
  Cavazos, and Cristina Silvano.
\newblock {COBAYN: Compiler Autotuning Framework Using Bayesian Networks}.
\newblock {\em TACO}, 13(2), 2016.

\bibitem{less-is-more}
Kyriakos Georgiou, Craig Blackmore, Samuel Xavier-de Souza, and Kerstin Eder.
\newblock Less is more: Exploiting the standard compiler optimization levels
  for better performance and energy consumption.
\newblock In {\em SCOPES}, 2018.

\bibitem{corl}
Rahim Mammadli, Ali Jannesari, and Felix Wolf.
\newblock {Static Neural Compiler Optimization via Deep Reinforcement
  Learning}.
\newblock In {\em LLVM-HPC}, 2020.

\bibitem{polygym}
Alexander Brauckmann, Andr{\'e}s Goens, and Jeronimo Castrillon.
\newblock {A Reinforcement Learning Environment for Polyhedral Optimizations}.
\newblock {\em arXiv:2104.13732}, 2021.

\bibitem{deeptune}
Chris Cummins, Pavlos Petoumenos, Zheng Wang, and Hugh Leather.
\newblock {End-to-end Deep Learning of Optimization Heuristics}.
\newblock In {\em PACT}, 2017.

\bibitem{loop-learner}
Rahim Mammadli, Marija Selakovic, Felix Wolf, and Michael Pradel.
\newblock {Learning to Make Compiler Optimizations More Effective}.
\newblock In {\em MAPS}, 2021.

\end{thebibliography}

\end{document}